# Design of a Modular GaN-based Three-Phase Three-Level ANPC Inverter

Angelo Di Cataldo
DIEEI
University of Catania
Catania, Italy
angelo.dicataldo@phd.unict.it

Hamed Eivazi
DIEEI
University of Catania
Catania, Italy
hamed.eivazi@unict.it

Giuseppe Aiello
SRA
STMicroelectronics
Catania, Italy
giuseppe.aiello01@st.com

Dario Patti
SRA
STMicroelectronics
Catania, Italy
dario.patti@st.com

Giacomo Scelba
DIEEI
University of Catania
Catania, Italy
giacomo.scelba@unict.it

Mario Cacciato
DIEEI
University of Catania
Catania, Italy
mario.cacciato@unict.it

Francesco Gennaro
SRA
STMicroelectronics
Catania, Italy
francesco.gennaro@st.com

*Abstract*— This paper presents the design of an 800 V 11 kVA three-level three-phase active neutral point clamped inverter, utilizing 650 V gallium nitride enhancement-mode high-electron-mobility transistors, to evaluate its feasibility in electric traction systems. The modular approach of the presented power converter design is detailed discussed and the different printed circuit boards composing the power converter are presented, together with critical design issues. The paper includes the gate driver design, as well as the thermal analysis and parasitics extraction using ANSYS Q3D Extractor.

*Keywords*— Multilevel inverters, gallium nitride devices, design optimization, gate driver, thermal analysis, parameter extraction.

I. INTRODUCTION

Nowadays, the inadequate range of electric powertrains is one of the main reasons hindering the widespread adoption of electric vehicles (EVs). Long battery charging times compared to the equivalent category of internal combustion vehicles is a further limitation together with poor battery autonomy generally less than 450km as optimistic estimation. Reduction in charging times could be a good compensation to autonomy issue on the consumer point of view. A higher charging power can reduce the battery refueling time and for this reason EV manufacturers are currently working at 800V-battery pack solutions, since an increase in DC current range would bring to higher losses [1-3].
At the same time, traction electric drives are required to feature high efficiency, compactness, high power density, high reliability and low weight. Gallium Nitride (GaN) technology presents a promising opportunity to achieve this target; the most outstanding feature of GaN enhancement-mode high-electron-mobility transistors (HEMTs) is their capability to switch at high frequencies improving the electromagnetic, electrodynamic and thermal behaviour of the motor downstream, simplifying the power converter's output filter and thus reducing the total size of the power conversion system [4,5]. Due to GaN HEMTs' relatively low maximum drain-to-source breakdown voltage (650 V), for high DC-bus voltage-electric traction applications this technology cannot be easily combined with two-level power inverters traditionally employed inside vehicles' powertrains [6]. For this reason, GaN-based multilevel inverter topologies are considered an effective technical solution for the next generation of electric traction units operating at high DC voltages, as they allow to reduce the voltage stress on switching power devices thus overcoming the breakdown voltage limitation of GaN technology. In addition, multilevel inverter topologies can be considered to gain lower current and voltage total harmonic distortions, lower electromagnetic interference and common mode voltages, higher efficiency, reliability, and power density [7,8]. Among multilevel inverter topologies, Active Neutral Point Clamped (ANPC), which is shown in Fig. 1, represents a good compromise between performances, compactness, efficiency and investment cost, and it is one of the most competitive solutions, as it allows to overcome the unequal power loss distribution among semiconductor devices shown by other multilevel converters, thus, enabling a substantially improved performance for high-power electrical drives [9-11].
In literature several designs for three-level GaN-based ANPC converter have been presented [12-15]. In [12,13] GaN power modules are proposed for a half-bridge three-level ANPC inverter, where power, driving, and conditioning circuits are integrated on the same PCB. In [12], an eight-layers printed circuit board (PCB) is used, and GaN HEMTs are paralleled operated according to the chosen modulation strategy, with three devices operating at high switching frequency and two devices switching at low-frequency. In [13], a four-layer PCB is employed, with GaN HEMTs placed on the top and bottom layers, and their switching pattern is controlled via optical receivers. In [14], a half-bridge three-level ANPC inverter is presented with an insulated metal core substrate (IMS) power section for bottom-side cooled GaN HEMTs, accompanied by two other PCBs for driving and power supply with a vertical totem stack configuration. In [15], a three-phase three-level ANPC inverter is proposed, where power, driving and conditioning circuits are placed on a two-layer PCB. GaN HEMTs' switching is controlled via STM32 microcontroller which can be plugged into the board.
This paper introduces a modular three-phase three-level ANPC inverter. It consists of:

- A main board housing power supply, DC-side and AC-side voltage and current sensing; three single-phase power cells consisting of six GaN HEMTs, decoupling capacitors and RC snubbers to mitigate drain-to-source overvoltages;
- A driving daughterboard with isolated gate drivers and a dual output DC/DC stage for each device, eliminated the need for any additional conditioning circuit;
- Two control boards are included, enabling the alternative control of switching devices with either optical fibers or STM32, respectively.

The proposed solution allows for the decoupling of power and signal parts of the converter, ensuring extreme modularity in the control approach, the employed power devices, and even the converter topology. Change can be made by simply acting on the driving and on the power cells. Moreover, the design stage includes thermal analysis and parasitics evaluation of the designed board, the latter accomplished by using the Ansys Q3D Extractor tool. The main board also includes a connector that allows to combine the inverter with an output low-pass filter directly on the same PCB. High switching frequencies can indeed negatively affect the electrical machine lifetime by contributing to insulation deterioration, leading to motor failure. However, high switching frequencies also enable the design of small-size output filters. Consequently, it becomes feasible to implement a passive filter directly into the GaN inverter, eliminating the need for a bulky and costly filter inductor [16].

The remaining part of the paper is organized as follows. Section II analyses the topology and the modulation strategies that were considered in the implementation of the ANPC inverter. In Section III the converter's layout is presented, describing all the boards composing the power conversion unit. Section IV illustrates the thermal analysis for a single-phase power cell of the inverter prototype, utilizing an electrical thermal model. In Section V, an estimation of power loop parasitic inductances for the GaN daughterboard is presented, conducted through a finite-element analysis (FEA) using Ansys Q3D Extractor software. Finally, section VI provides concluding remarks.

## II. INVERTER TOPOLOGY AND MODULATION STRATEGIES

Three-phase three-level ANPC inverter topology is displayed in Fig. 1. The name comes from the two 'active' devices, unlike the NPC in which two diodes are used. These active devices are exploited to clamp the output voltage to the neutral point of the DC-link when the output zero-voltage level is required. It consists of six switches per phase that can be commutated according to sinusoidal pulse width (SPWM) or space vector (SVPWM) modulation strategies, with the main aim to minimize conduction and switching power losses [17]. Several multicarrier strategies have been developed for multilevel inverters to reduce the output voltage distortion. Some methods use carrier disposition, as in the Phase-Disposition (PD) strategy, while others use phase shifting of multiple carrier signals [18]. In literature some PD multicarrier PWM strategies have been specifically proposed for ANPC, as the Diode Neutral Point Clamped modulation (DNPC), the ANPC modulation with Same-Side Clamping (ANPC-SSCM), the ANPC modulation with Opposite-Side Clamping (ANPC-OSCM), and the ANPC with Full-Path Clamping (ANPC-FPCM). According to [11], the ANPC-FPCM allows to get lower total losses, whereas in [19], the ANPC-OSCM is revealed as the best strategy in terms of common-mode noise reduction. In SVPWM the switches' gating pulses are generated by using the concept of space vectors starting from the 27 feasible switching states of a three-level inverter. For the same modulation index, SVPWM produces lower harmonic distortion than SPWM, and it allows to better utilization of the DC-bus [20]. The presented layout facilitates the implementation of all the aforementioned modulation techniques.

## III. INVERTER TOPOLOGY AND MODULATION STRATEGIES

The proposed design consists of four main boards corresponding to the four different sections that make up the modular three-phase three-level ANPC inverter:

    A. Main Board;

    B. Three GaN Daughterboards (GaN DBs);

    C. Three Driving Daughterboards (Driving DBs);

    D. Control Daughterboard (Control DB).

Below the description of each boar is provided.

### A. Main Board

Main board is a 4-layers PCB which is made of the DC-side, DC and AC current and voltage sensing, power connections, and all the connectors that allow to easily connect the daughterboards. Fig. 2 shows the 3D top view of the main board with the daughterboards connected above. The main board's design is carried out with the main goal to realize a flexible power PCB. In addition, this modularity allows to sunder power and signal sections, reducing interferences and optimizing the overall size of the power converter. Main board's DC-side is made up of input connectors for supplying both power and signal sections, bulk (DC-link) and decoupling capacitors, and paralleled resistors to guarantee an uniform voltage distribution to both pairs of bulk capacitors. Bulk capacitors are in charge to keep stable the DC input voltage. Therefore, four electrolytic capacitors are chosen due to their highest capacitance-to volume ratio, making them ideal for reduction of DC-bus voltage ripple. Capacitance value $C_{DC}$ is chosen according to (1) to keep the oscillations within the limits of ±5%, where $V_{pp}$ is the peak-to-peak DC voltage, $V_{DC}$ is the nominal DC voltage (targeted to 800 V), $\omega$ is the output voltage angular frequency (set equal to 314 rad/s) and $I_{max}$ is the maximum value of the output current.

$$C_{DC} = I_{max} \frac{V_{DC}}{4 V_{pp}^2 \omega} (1 - (cos^2(2 \omega p))) \qquad (1)$$

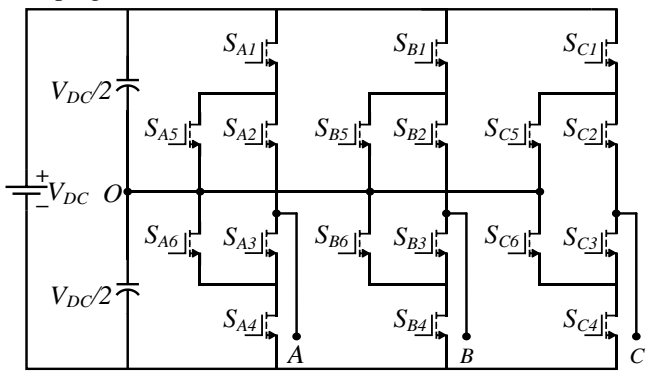

Fig. 1. Three-phase three-level ANPC inverter topology.

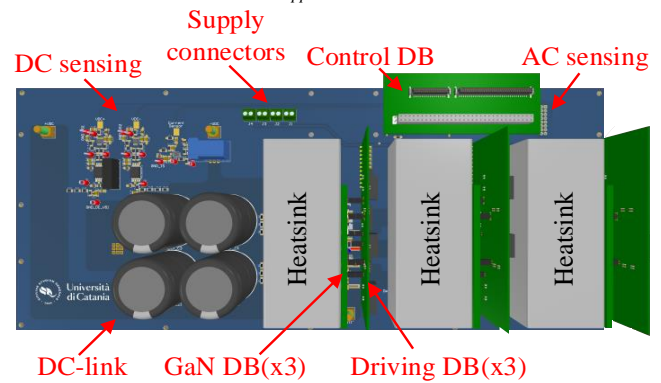

Fig. 2. 3D top view of assembled main board.

Nevertheless, electrolytics have considerably high equivalent series resistance (ESR) and equivalent series inductance (ESL). Parasitic power loop inductance affects the switching losses especially at high frequency operation [21]. Therefore, six low ESR and ESL multi-layers ceramic capacitors (MLCCs) are paralleled to the bulk capacitors to counteract the undesired effects of ESR and ESL. They are positioned as close as possible to the connectors of the GaN daughterboards to minimize the impact of stray inductance on transient commutations.

Sensing circuits are responsible for sensing the DC and AC voltages and currents, enabling not only the measurement of these electrical quantities but also providing feedback analog signals to the control board. This allows the implementation of control strategies without the need for any any external circuitry, a crucial aspect in reducing the overall size of the electric drive.

### B. GaN Daughterboard

GaN daughterboard (Fig. 3) consists of a single-phase power cell, where the six GaN HEMTs, decoupling capacitors and RC snubbers are located.

GaN DB is connected to the main board through press-fit power connectors that mechanically support the PCB. The connectors-to-main board must be dimensioned according to the rating of the RMS output phase current $I_{out(RMS)}$ which is expected to flow through each of them. The computation of $I_{out(RMS)}$ is conducted according to (2), assuming sinusoidal output current waveforms.

$$I_{out(RMS)} = \frac{2\sqrt{2}\, P_{out}}{m_a\, V_{DC}} \quad (2)$$

Considering a target output power $P_{out}$=11 kW, a DC-bus voltage $V_{DC}$=800 V and an amplitude modulation index $m_a$=0.8, the value of $I_{out(RMS)}$ is 16 A.

In the GaN daughterboard, the need for additional decoupling capacitors is crucial to minimize the power loop inductance in proximity of switching devices. Therefore, five paralleled 200 nF MLCC capacitors are chosen for both the positive and negative halves of the DC-link. The board employs SGT65R65AL e-mode GaN HEMTs (Tab.1) manufactured by STMicroelectronics. Using bottom-side cooled devices allows the placement of surface mount technology (SMT) decoupling capacitors close to the device, as the heatsink is mounted on the other external layer of the PCB. This GaN daughterboard exploits IMS technology which offers lower case-to-heatsink thermal resistance to maximize the heat dissipation of the bottom-side cooled devices [22]. Despite this, IMS solution limits the layout design to a single-layer PCB, preventing the use of through-hole components and vias, which are often useful for minimizing conductive paths.

TABLE I. MAIN SPECIFICATIONS OF SGT65R65AL.

| Parameter | Value |
| --- | --- |
| Drain-to-source blocking voltage | 650 V |
| Drain-to-source blocking voltage (transient<1µs) | 750 V |
| Gate-to-kelvin source voltage | -10 V to 7 V |
| Drain current (continuous) | 25 A |
| Typical static drain-source on-resistance | 49 mΩ |
| Maximum static-source on-resistance | 65 mΩ |
| Operating junction temperature | -55°C to +150°C |
| Turn-on switching energy | 33.8 µJ |
| Turn-off switching energy | 19.5 µJ |

GaN HEMTs allow to work at desirable high switching frequencies, leading to high *dv/dt* and *di/dt*. In particular, the latter can cause high overshoots in drain-to-source voltages. For this reason, RC dissipative snubbers are paralleled to each device to smooth transient overvoltages. Although this results in the dissipation of energy and slower commutations, the value of dissipated energy is low, considering the low intrinsic output capacitance of GaN HEMTs. The benefit of reduced overvoltages brings a substantial advantage in placing snubbers [23].

GaN HEMTs are connected to driving DB by means of board-to-board connectors, linking gate drivers' outputs with the respective gate and Kelvin source pins of GaN HEMTs. The separation of power cell and driving circuits prevents perturbations on the gate-to-kelvin source voltages caused by high current flowing in overlapped conductive paths, even if in different layers of the same PCB, due to crosstalk.

### C. Driving Daughterboard

The four-layers driving daughterboard (Fig. 4) consists of a DC-DC stage and a driving circuit for each GaN HEMT of the ANPC leg. It is orthogonally connected to the main board by means of a connector which provides supply voltages (5 V for DC-DC converter, 3.3 V for Gate Driver and 12 V for fans placed in the GaN daughterboard) and driving signals from and to the main board, respectively. It is mechanically paralleled to the GaN daughterboard, and connected to it by means of transversal connectors that provide gate and Kelvin source signals, as explained in Section III B.
The DC-DC stage consists of MGN1D050603MC-R7 dual output 1 W isolated DC-DC converters, which convert the 5 V input voltage (properly smoothed by means of a by-pass capacitor) to 6 V/-3 V output voltage levels. The use of a dual DC-DC converter allows to significantly reduce the size of the DC-DC stage since no conditioning circuit is required. DC-DC converter's output voltage levels are exploited as output reference voltages of the STGAP2GSNCTR galvanically isolated gate driver manufactured by STMicroelectronics for the GaN HEMTs piloting. The high voltage isolation between input and output pins prevents electric arcs and short circuit between input (low voltage, LV) and output (high voltage, HV) sides of the component. The gate driver is the core element of the driving circuit, whose schematic is illustrated in Fig. 5. The PWM signal of each device is provided in input to the gate driver ($IN+$ and its ground $IN-$), together with the 3.3 V supply voltage ($V_{DD}$ and its ground $GND$) which is smoothed by two input by-pass capacitors, $C_1$ and $C_2$. Supply and PWM grounds are short circuited. In addition, a RC filter is placed to eliminate high-frequency noise. For the filter design, after the resistance $R_f$ value is set equal to 1 kΩ, and the cutoff frequency $f_{cut}$ is imposed to two orders of magnitude greater than the predicted switching frequency (100 kHz), the capacitance value $C_f$ is set to 16 pF according to the well-known relation expressed in (3).

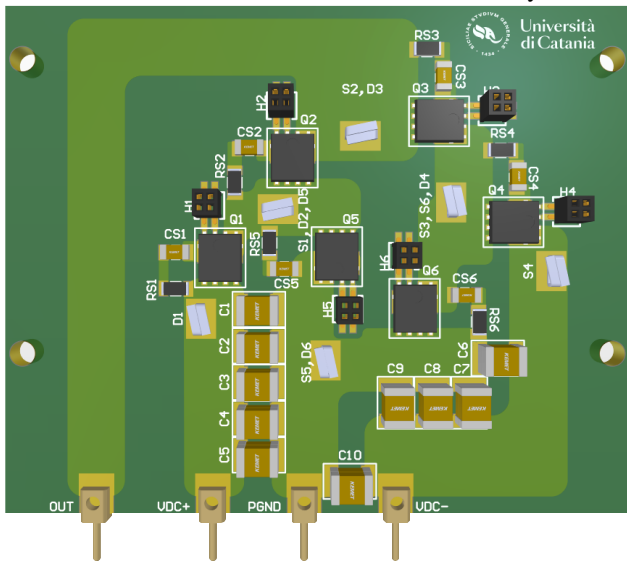

Fig. 3. 3D top view of GaN daughterboard.

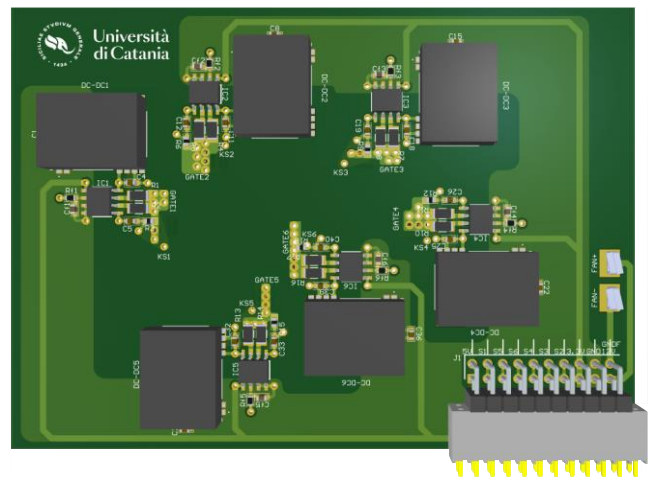

Fig. 4. 3D top view of driving daughterboard.

$$C_f = \frac{1}{2\pi f_{cut} R_f} \quad (3)$$

In the HV-side of the gate driver two by-pass capacitors ($C_3$ and $C_4$) are symmetrically placed as the input ones and connected between $GND_{ISO}$ and $V_H$ pins. In addition, two other by-pass capacitors ($C_5$ and $C_6$) are placed between the kelvin source isolated ground of the device and the two DC-DC output reference voltages for turning-on and turning-off, which are 6 V and -3 V, respectively. 6 V is the optimal on-state gate voltage for SGT65R65AL, whereas the negative bias (-3 V) for the turning-off voltage increases immunity to spurious gate bouncing and false switch turn-on. Consequently, it is the best configuration for hard-switching high-power applications, when high $dv/dt$ occurs during switching transients [24]. Finally, the gate resistances $R_{g(on)}$ and $R_{g(off)}$ are important to set how fast the device goes through the turning-on and in the turning-off, respectively. They are connected to $G_{ON}$ and $G_{OFF}$ pins, and their values are determined according to the inequalities in (4) and (5):

$$R_{g(on)} \geq \left(\frac{V_{CC} - GNDA}{I_{OH}} - R_{DS(on)}\right) \quad (4)$$

$$R_{g(off)} \geq \left(\frac{V_{CC} - GNDA}{I_{OL}} - R_{DS(on)}\right) \quad (5)$$

where $V_{CC-GNDA}$ is the voltage difference between the two output voltages values provided to the gate driver's output pins, thus 9 V. $R_{DS(on)}$ is the typical static drain-source on-resistance, while $I_{OH}$ and $I_{OL}$ are the maximum allowed output currents during the turning on and the turning off transients, which for the employed gate driver are equal to 2.5 A and 3.75 A, respectively. $R_{g(on)}$ is set to 10 Ω, whereas $R_{g(off)}$ is set to 2.2 Ω. In addition, a large resistor $R_1$ is added to pull-down to 0 V the gate-to-kelvin source voltage when the input signal is open circuited, and thus the gate terminal is floating. The proposed daughterboard's layout considers the optimization of the gate driver loop since it is one of the most crucial design aspects to deal with when high $dv/dt$ and $di/dt$ occur. Because of the common source inductance/mutual inductance, the $di/dt$ of the power loop can easily affect the gate-to-kelvin source voltage during the switching. On the other hand, the $dv/dt$ could result in Miller effect. SMT components are used, and conductive traces' lengths are minimized to achieve low gate loop parasitic inductance. In addition, 0603 packages are used for layout optimization purposes, reducing the board's overall size. Moreover, one of the PCB's inner layers is exploited to extend the LV signals' ground plane to guarantee a return path for LV signals in an adjacent layer with respect to the one in which they are routed. This creates a vertical flux cancellation that reduces the gate loop stray inductance.

### D. Control Daughterboard

The control daughterboard is responsible for providing PWM signals to gate drivers. Two alternative four-layers versions of this DB are designed to exploit either optical receivers or STM32 microcontroller. Both PCBs are compatible to be plugged into a main board's connector.
Optic control daughterboard, Fig. 6, consists of 18 optical receivers (one for each device) and their respective by-pass capacitors. This daughterboard allows to control the power converters' switches with Field Programmable Gate Array (FPGA), dSpace or microcontroller. Due to the weight of the transmitter-to-receiver cables, the board is mechanically supported by four screws that connect it to the main board.
STM32-adapter daughterboard, Fig. 7, consists of two connectors to plug-in the analog-32pins and the digital-68pins of a control board for STM32-G474QET6 Mainstream ARM Cortex-M4 microcontroller. This PCB is significantly smaller than the optics control board and it does not require screws for mechanical support. In addition to provide PWM signals, this board allows to implement closed-loop control strategies, since AC and DC sensing output signals are routed to microcontroller analog-to-digital converters (ADC) pins. However, this DB allows to implement three of the four SPWM strategies mentioned in Section II, that is the DNPC, the ANPC-SSCM and the ANPC-OSCM. Both ANPC-FPCM and SVPWM can be implemented only with optic control daughterboard.

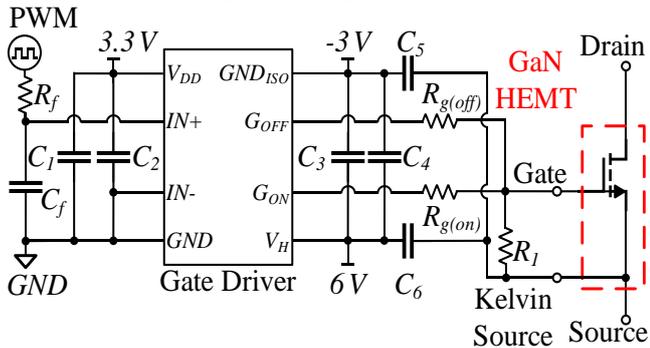

Fig. 5. Schematic representation of the gate driving circuit.

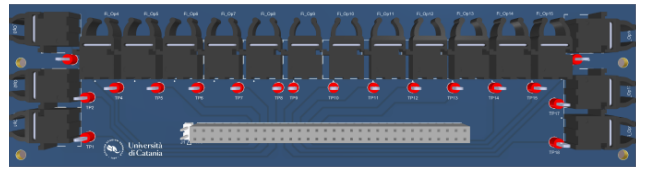

Fig. 6. 3D top view of optic control daughterboard.

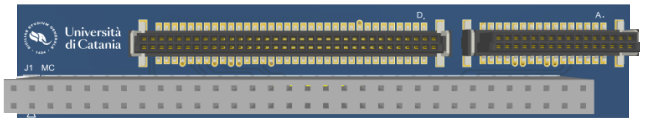

Fig. 7. 3D top view of STM32-adapter daughterboard.

## IV. THERMAL ANALYSIS

Thermal analysis is fundamental to forecast the heat dissipation requirements and to correctly dimension the conductive traces of the PCB. Thermal issues are more relevant in the GaN daughterboard, on which a thermal analysis is conducted. The aim of the thermal analysis is to focus on steady-state conditions, thus stray capacitances are not considered in the present model, as in [25]. The different layers involved in the heat extraction are illustrated in Fig. 8.a from the single bottom-side cooled GaN HEMT to the external heatsink placed in addition to IMS. A thermal interface material (TIM) is required between IMS and heatsink to prevent electrical short circuit among GaN HEMTs' thermal pads, which are connected to respective dies' source pins inside the device's package.

The single GaN HEMT's heat flux to be dissipated is equal to the sum of the device's conduction $P_{cond}$ and switching $P_{sw}$ power losses. For the thermal analysis the switching frequency $f_{sw}$ is set equal to 100 kHz, the DC voltage equal to 800 V and the RMS value of the output phase current equal to 16 A, as get from (2). For conduction losses the GaN HEMT is assumed to be always in the on-state for sake of simplicity. Considering the typical static drain-source on-resistance $R_{DS(on)}$ and the energies dissipated during the turn-on ($E_{on}$) and turn-off ($E_{off}$) transitions reported in Tab.I, the overall losses $P_{tot}$ are determined as:

$$P_{tot} = P_{cond} + P_{sw} = R_{DS(on)} I_{out(RMS)}^2 + (E_{on} + E_{off})f_{sw} \quad (6)$$

GaN HEMTs dissipate heat in different ways according to their behaviour connected to the chosen modulation technique. For sake of simplicity, they can be assumed as equal heat sources. The resulting equivalent thermal circuit is shown in Fig. 8.b, in which $Q$ is the heat flux to be dissipated by each GaN HEMT (equal to $P_{tot}$), while the other parameters are specified in Tab. II. $R_{sa}$, is the heatsink-to-ambient thermal resistance which can be evaluated according to (7), where $n$ is the number of devices to be cooled by the system (equal to 6).

$$R_{sa} = \frac{T_j - T_a}{n P_{tot}} - R_{TIM} - R_{IMS} - R_{FR4} - R_{Cu} - R_{sm} - \frac{R_{jc}}{n} \quad (7)$$

TABLE II. PARAMETERS OF THERMAL ANALYSIS.

| Description | Symbol | Value |
|---|---|---|
| Junction-to-case thermal resistance | $R_{jc}$ | 0.41°C /W |
| Copper layer thermal resistance | $R_{Cu}$ | 4e-5°C /W |
| FR4 thermal resistance | $R_{FR4}$ | 0.06°C /W |
| IMS thermal resistance | $R_{IMS}$ | 0.32°C /W |
| TIM thermal resistance | $R_{TIM}$ | 0.08°C /W |
| Junction temperature | $T_j$ | 130°C |
| Ambient temperature | $T_a$ | 25°C |

A LAM47512 heatsink manufactured by Fischer Elektronik fits the requirements both in dimensions and in thermal resistance, which is comprised between 0.6°C/W and 1.1°C/W. In addition, it presents an integrated 12 V-supply fan that boosts the cooling of the system.

## V. PARASITICS ESTIMATION

Minimization of total commutation inductance is crucial to reduce voltage overshoots and switching losses, as explained in Section III.B. It is formed by the intrinsic inductance of the DC-link and the power loop parasitic inductance. The former can be minimized paralleling low-ESL decoupling capacitors, the latter acting on the layout. For this reason, a parasitic estimation is needed in the design stage to optimize the layout of the GaN DB in which commutation loop is mainly contained. The overall power loop parasitic inductance can be split in single contributes relative to the traces that make it up. Fig. 9.a shows the electrical circuit comprehensive of the single inductive contributes associated to the layout traces illustrated in Fig. 9.b. Each stray inductive contribute $L_\sigma$ can be analytically obtained by:

$$L_\sigma = \frac{\mu_r \mu_0 h l}{w} \quad (8)$$

where $\mu_r$ is the FR4 relative permeability, $\mu_0$ is the air permeability, $h$ is the copper layer thickness, $l$ and $w$ are the length and the width of the considered trace, respectively. In addition to the analytical approach, parasitic inductances were estimated by means of a FEA in Ansys Q3D Extractor tool. Parasitic inductances behavior was observed in a large frequency spectrum, firstly conducting a DC analysis and then an AC one at 100 kHz (equal to $f_{sw}$). In both cases, the presence of real devices is neglected, considering only the copper traces. Stray resistances have negligible values, while stray inductances' values obtained at 100 kHz are reported in

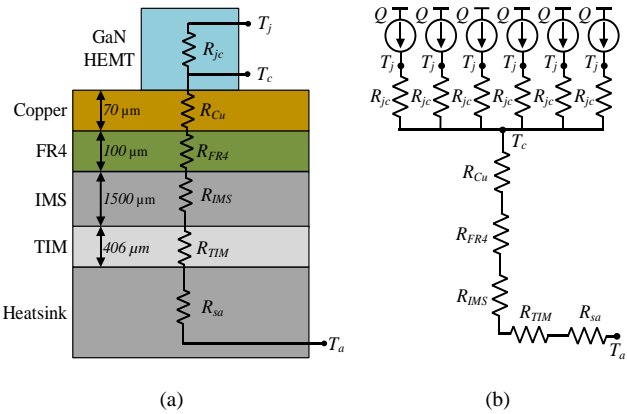

Fig. 8. Scheme of thermal layers (a) involved in GaN daughterboard's heat dissipation and relative equivalent thermal circuit for heat dissipation of six devices (b).

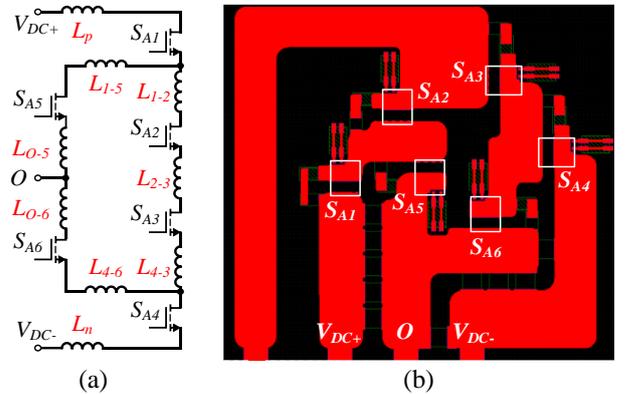

Fig. 9. Electrical circuit (a) comprehensive of stray inductances related to GaN daughterboard's designed traces (b).

TABLE III. PARASITIC INDUCTANCES ESTIMATED WITH ANSYS Q3D EXTRACTOR.

| Inductance | Value@100kHz | Inductance | Value@100kHz |
|---|---|---|---|
| $L_p$ | 14.46 nH | $L_n$ | 22.04 nH |
| $L_{1-2}$ | 5.08 nH | $L_{4-3}$ | 3.78 nH |
| $L_{1-5}$ | 5.72 nH | $L_{4-6}$ | 6.15 nH |
| $L_{O-5}$ | 13.93 nH | $L_{O-6}$ | 11.04 nH |
| $L_{2-3}$ | 4.84 nH | | |

Tab. III. Finite-element analysis's results show low parasitic inductances' values, and demonstrate good layout simmetry: $L_{1-2}$ is almost equal to $L_{1-5}$, and the same happens for $L_{4-3}$ and $L_{4-6}$, and for $L_{O-5}$ and $L_{O-6}$. The power traces which come from connectors-to-main board ($L_p$, $L_n$, $L_{O-5}$ and $L_{O-6}$) have higher values, as expected considering the length of the traces shown in Fig. 8.b, and according to (8). However, these stray inductances' impact is smoothed by the presence of the decoupling capacitors discussed in Section III.B.

VI. ACKNOWLEDGMENT

This work was carried out within the ECSEL-JU project GaN4AP (GaN for Advanced Power Applications), under grant agreement no. 101007310. This Joint Undertaking receives support from the European Union's Horizon 2020 research and innovation programme.

VII. CONCLUSIONS

The presented design of a three-phase three-level ANPC converter offers as main benefit the modularity. This feature makes the converter flexible to be exploited for different topologies limiting future changes or improvements to the design of small daughterboards, saving costs and time. In addition, the modular structure of the power conversion unit allows to sunder different sections of a power converter as DC-link, driving circuits and commutation loops whose overlapping in the same PCB could bring to disturbances, overhoots and increased losses. Finally, a preliminary evaluation of the proposed design is validated through the parasitics estimation, confirming low stray inductances' values of the traces that are designed in the switching power cell. However, these preliminary simulations can be improved by considering the presence of real devices on the PCB.